\documentclass[usenatbib, onecolumn]{mnras}
\usepackage[utf8]{inputenc}
\usepackage{fix-cm}
\usepackage{amsmath, amssymb, graphicx, booktabs, physics} 
\usepackage{lipsum}
\usepackage{caption}

\title{Harmony of the Spheres: Extension to All Points of an Algorithm for Producing a Density Field with Given Two-, Three-, and Four-Point Correlation Functions}

\author[Z. Slepian \& A. Greco]{Zachary Slepian$^{1,2}$ \& 
Alessandro Greco$^{1}$\\
$^{1}$Department of Astronomy, University of Florida, 211 Bryant Space Science Center, Gainesville, FL 32611, USA\\
$^{2}$Lawrence Berkeley National Laboratory, 1 Cyclotron Road, Berkeley, CA 94720, USA}

\date{}

\begin{document}
\label{firstpage}
\pagerange{\pageref{firstpage}--\pageref{lastpage}}
\maketitle

\begin{abstract}
Previous work \citep{Slepian:2023gni} showed that the \cite{smith_zald_2011} algorithm to realize Cosmic Microwave Background (CMB) maps with any desired harmonic-space bispectrum could be generalized to produce a 3D density field with any desired N-Point Correlation Functions  (NPCFs, N = $2, 3, \ldots$) about a particular, specified set of ``primary'' points. This algorithm assured one of having the correct correlations if measured about these specific centers. Here, we show that this algorithm was more general than initially believed, and can in fact be used to produce a density field on a grid that has the correct, desired NPCFs as measured about \textit{every} point on the grid. This paper should be considered the second in the series, and now completes the quest to generalize the idea of ``constrained realization'' \citep{hoffman_ribak} to higher-order statistics. This algorithm will be of great use for quickly generating density fields both to produce covariance matrices, and test systematics, for current and future 3D large-scale structure surveys such as Dark Energy Spectroscopic Instrument (DESI), Euclid, Spherex, and Roman.
\end{abstract}


\section{\label{sec:introduction}Introduction}
Recently, \cite{Slepian:2023gni} showed that perturbing the spherical harmonic coefficients of a density field on shells around a given set of points could be used to produce any desired NPCF about those points, writing down explicit prescriptions for $N = 2 - 4$ and showing the method generalizes to $N > 4$ as well (see \citealt{encyc} for a recent review article on 2 and 3-point statistics and beyond). This algorithm was an expansion on the idea of \cite{smith_zald_2011} to accomplish the same task for the Cosmic Microwave Background (CMB) angular bispectrum, where one only needs to do it about one point (the observer, us) and, roughly on one shell (at recombination).\footnote{In full detail, their algorithm produces the desired $a_{\ell m}$ given the angular bispectrum as an input. These $a_{\ell m}$ do not correspond just to the shell of recombination, but they are those obtained summing all the shells, in a line-of-sight integral. Of course, since this integral is weighted by the visibility function, the main contribution will come from recombination, but in principle the algorithm captures the whole history, including \textit{e.g.} reionization.}

\cite{Slepian:2023gni} stated that the algorithm there presented only led to the desired NPCFs about the set of ``primary'' points about which the shells were created under a certain restriction. One had to ensure that each point was more distant from all others than double the maximum scale out to which correlations are measured, $R_{\rm max}$. That is, one needed to have some set of primary points each separated from each other by at least $2R_{\rm max}$, so that the shells about one point did not ``interfere'' with shells about any other point. This situation is shown in Fig. \ref{fig:two_spheres_no_overlap}.


This was a significant restriction---it meant that one had to instruct the algorithm used to measure correlations (\textit{e.g.} \textsc{ENCORE}, \citealt{encore} or \textsc{SARABANDE}, \citealt{sunseri}) to compute correlations only about those points. Further, it meant that one was limited in the number of primary points one could use in a given volume. This limit was roughly $V/V_{\rm max}$, with the latter being the volume of a sphere with radius $R_{\rm max}$.

In short, the method of the previous work seemed to be not quite a way to produce a \textit{full} density field with a given set of correlations, but rather, a density field where, around some set of \textit{special points}, one had the desired correlations.

This would still be useful for testing localized systematics, such as fiber collisions in a galaxy redshift survey like Sloan Digital Sky Survey Baryon Oscillation Spectroscopic Survey (SDSS BOSS, \citealt{hahn_fiber}) or Dark Energy Spectroscopic Instrument (DESI), \textit{e.g.} \cite{ pinol, burden, hand_pk, bianchi, pinon}. These affect only small scales and could be treated as uncorrelated on larger scales; thus, one would not need to track cross correlations between primary points. 

However, for general applications, this apparent property of the algorithm could still seem a limitation. In particular, if one wished to use such realizations for covariance matrices (\textit{e.g.} realize a given set of 2, 3 and 4-point statistics to make boxes to assess 2-point function covariance, or a given set of 2, 3, 4, 5, and 6-point statistics to assess 3-point function covariance) this limitation would be fatal. Computing a covariance from a set of realizations would require that the correlations in the realizations were correct about all points, not just a specified set, as the covariance is a set of integrals over all space \textit{e.g.} \cite{xu_2012} (2PCF) , \cite{3pt_alg} (3PCF), and \cite{hou_covar} (4PCF and beyond).

While as surveys like DESI continue to future years and more passes over each part of the sky and thus higher completeness, the fiber collision effect will diminish in importance, the higher precision of these increasing volumes will place ever-more stringent demands on the precision of the covariance. This will be true for model-dependent searches such as standard analyses of the 2 and 3-point statistics, and especially important for model-independent searches such as for parity violation (PV) (\textit{e.g.} \citealt{cahn_parity_prl}), which have thousands to tens of thousands of degrees of freedom \citep{hou_parity, phil_parity, slepian_parity} and thus require a number of mocks well beyond the reach of conventional mock-generation methods such as N-body (\textit{e.g.} \texttt{Abacus}, \citealt{abacus}) or approximated dynamics (\textit{e.g.} \texttt{EZmocks}, \citealt{ez-mocks}). 

PV searches in particular are very sensitive to precise estimate of the covariance matrix, as discussed in \cite{cahn_parity_prl} and highlighted further in \cite{krolewski}. Having a fast method to produce fields with a given set of NPCFs would offer a path to thousands of mocks that could be used for more precise covariance matrices and address this serious challenge faced by model-independent PV searches. In detail, one could imagine measuring the NPCFs on the data up to some order (say $N=6$, the current maximum of \textsc{ENCORE}) and then producing thousands of fast mocks with matching NPCFs using the methods outlined in \cite{Slepian:2023gni} and in the present paper. This approach would guarantee (up to a given $N$ and up to cosmic variance) \textit{no} data-mock mismatch, an issue that \cite{krolewski} argues is central for PV studies. 

In the current work, we show that in fact the algorithm of \cite{Slepian:2023gni} can be used to produce a density field with the desired NPCFs about \textit{every} point of a grid. This occurs because the Gaussian Random Fields (GRFs) upon which the algorithm is built can all be independent. Thus, the ``interference'' of shells about one point with correlations about another will in fact vanish, even if the shells overlap (the situation shown in Fig. \ref{fig:two_spheres_overlap}). There is no need to have separated, non-intersecting regions. With this restriction removed, one can realize the desired correlations about every point on a given grid, to get a continuous density field with these correlations.\footnote{A continuous field is represented computationally as values at every point on a grid.} This then enables production of fast mocks that match the data up to a desirred NPCF, and can be used for covariances for both standard, model-dependent analyses and model-independent PV searches.

We perform a simple set of calculations to show how the cancellation of correlations between different shells works in practice. The underlying principle is very simple: if one makes all the fields statistically independent of each other, then all cross-correlations between fields around two or more different primary points vanish. The independent fields never ``see'' each other. 

With this work, then, we complete the task of obtaining an algorithm to give a density field that truly does have any desired NPCFs as measured about any point of that field.

The work is structured as follows. First, in \S\ref{sec:rvw} we briefly review the algorithm of \cite{Slepian:2023gni}. In \S\ref{sec:cross-correlation}, we compute the cross-correlation of the shells around two different primary points, and show that it vanishes. This argument extends to correlations of any larger number of primary points because these can always be split into pair-wise correlations by Wick's (Isserlis') theorem, as the underlying fields used are Gaussian Random Fields (GRFs). Next, we turn to the computational cost of our algorithm in \S\ref{sec:computational_cost}. We close in \S\ref{sec:conclusion}.

\section{Review of the algorithm for non-Gaussian density fields}
\label{sec:rvw}
We here briefly review how the \cite{Slepian:2023gni} algorithm works to produce a desired 4-Point Correlation Function (4PCF) about a given set of primaries. The algorithm works analogously to produce a given 3PCF, but we do not detail that; we do account for the 3PCF-producing-term when we perform the calculation of \S\ref{sec:cross-correlation}.

\begin{figure}
\centering
\includegraphics[width = .7\linewidth]{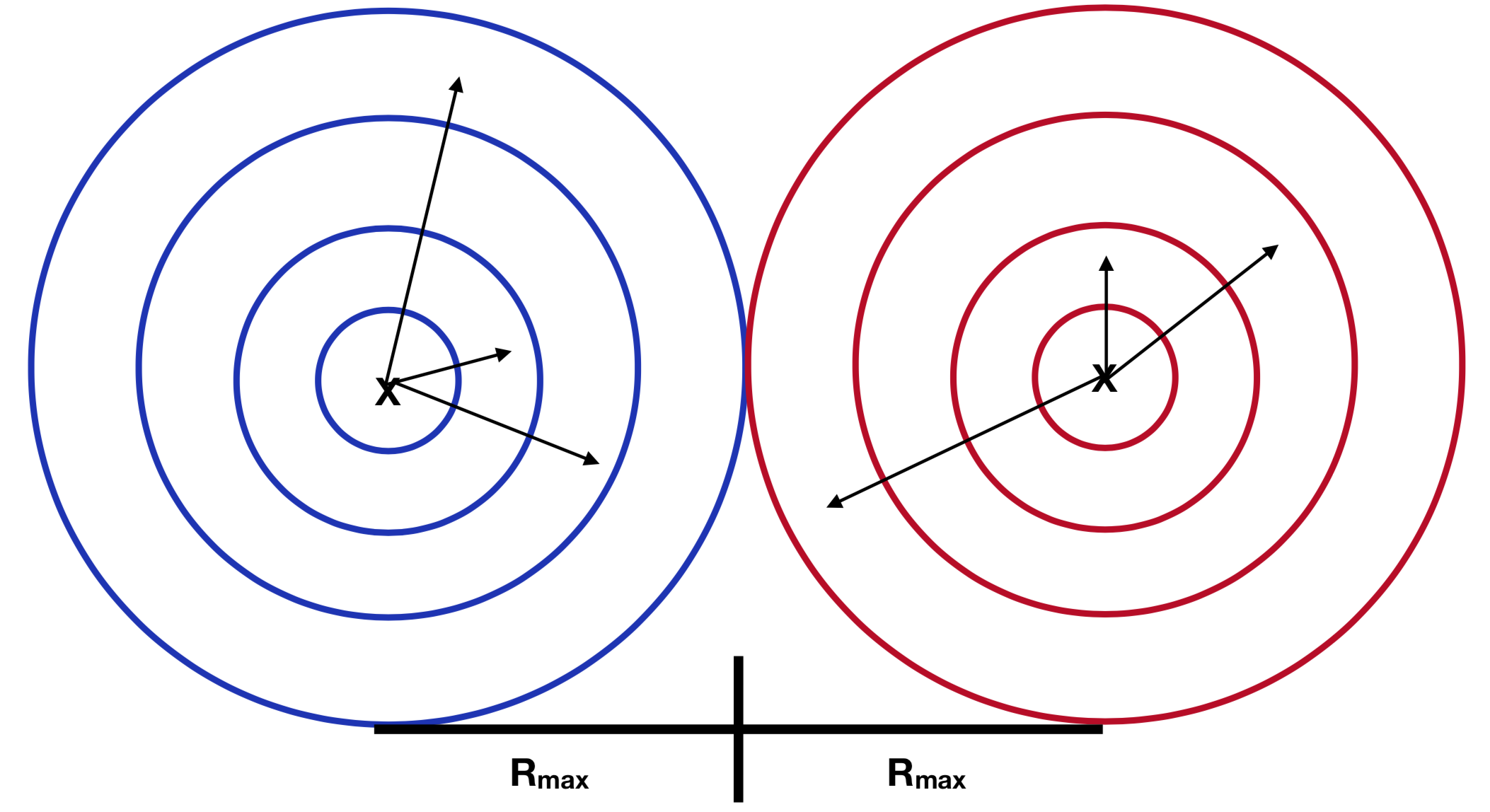}
\caption{Here we display two spheres, centered at $\vec{x}$ and $\vec{x}^{\,\prime}$, respectively (\textit{cf.} Eq. 1. At each center there is a primary galaxy, while the three secondary ones are located in the shells of radius $r_i$ at a given angular coordinate $\hat{r}_i$, here shown by arrows. The correlations are measured out to a radius $R_{\rm max}$, and here we show the configuration treated in Slepian (2024), where the two spheres are taken not to overlap, so that the correlations about one primary would not interfere with those about the second. In the current work, we show this restriction is unnecessary, and that the framework of Slepian (2024) also applies to the more general case, where the two spheres overlap, as shown in Fig. \ref{fig:two_spheres_overlap}.}
\label{fig:two_spheres_no_overlap}
\end{figure}

Let us first consider a ``primary" galaxy at the origin of a sphere with center $\vec{x}$, and three ``secondary" ones located on shells of radius $r_i$ at angular coordinates $\hat{r}_i$ (for instance, as shown on either the blue or the red side of Fig. \ref{fig:two_spheres_no_overlap}). The 4PCF, $\zeta$, of a density fluctuation field $\delta$ is an average over all primaries, as
\begin{equation}
\begin{split}
\label{eqn:4PCF}
\zeta(r_1,r_2,r_3;\hat{r}_1,\hat{r}_2,\hat{r}_3)&\equiv\langle\delta(\vec{x})\delta(r_1,\hat{r}_1;\vec{x})\delta(r_2,\hat{r}_2;\vec{x})\delta(r_3,\hat{r}_3;\vec{x})\rangle_{\vec{x}}\\
&=\sum_{\ell_1\ell_2\ell_3}\zeta_{\ell_1\ell_2\ell_3}(r_1,r_2,r_3)\mathcal{P}_{\ell_1\ell_2\ell_3}(\hat{r}_1,\hat{r}_2,\hat{r}_3).
\end{split}
\end{equation}
In the first line, the angle brackets represent averaging over $\vec{x}$ (\textit{i.e.} over all primaries). In the second line, $\zeta_{\ell_1\ell_2\ell_3}(r_1,r_2,r_3)$ are the coefficients of the expansion in the isotropic basis functions \citep{Cahn:2020axu}, which are
\begin{equation}
\label{eqn:isotropic_basis_function}
\mathcal{P}_{\ell_1\ell_2\ell_3}(\hat{r}_1,\hat{r}_2,\hat{r}_3)
\equiv (-1)^{\ell_1+\ell_2\ell_3}\sum_{m_1m_2m_3}\begin{pmatrix}
\ell_1 & \ell_2 & \ell_3 \\
m_1 & m_2 & m_3
\end{pmatrix}Y_{\ell_1m_1}(\hat{r}_1)Y_{\ell_2m_2}(\hat{r}_2)Y_{\ell_3m_3}(\hat{r}_3).
\end{equation}
We note that in the isotropic basis, we must have an ordering of the $\ell_i$ to eliminate the sign degeneracy under permutation in the 3-$j$ symbol when the sum of the $\ell_i$ is odd, as discussed more fully in \cite{Cahn:2020axu}. Following this latter, we choose  $r_1< r_2 < r_3$. Thus the three shells are distinguishable from each other.

We expand the density field $\delta$ on each shell (designated by $r_i$) about the primary galaxy at $\vec{x}$ as 
\begin{equation}
\label{eqn:density_field}
\delta(r_i,\hat{r}_i;\vec{x}) = \sum_{\ell = 0}^{\infty}\sum_{m=-\ell}^{\ell}a_{\ell m}(r_i;\vec{x})Y_{\ell m}(\hat{r}_i).
\end{equation}
The goal of the \cite{Slepian:2023gni} algorithm is to create a harmonic field on each shell (\textit{i.e.} an $a_{\ell m}$) that will give rise to the desired 4PCF when correlators of $\delta$ are computed. 

The algorithm first constructs the auxiliary quantity
\begin{equation}
\mathcal{T}^{(4)}[\vec{a}\,^{(4)}(\vec{x})]\equiv\sum_{r_1< r_2< r_3}\,\sum_{\ell_1\ell_2\ell_3}\,\sum_{m_1m_2m_3}\zeta_{\ell_1\ell_2\ell_3}(r_1,r_2,r_3)\begin{pmatrix}
\ell_1 & \ell_2 & \ell_3 \\
m_1 & m_2 & m_3
\end{pmatrix}a_{\ell_1m_1}(r_1;\vec{x})a_{\ell_2m_2}(r_2;\vec{x})a_{\ell_3m_3}(r_3;\vec{x})
\end{equation}
where we use vector notation to indicate a set of $a_{\ell m}$, as
\begin{equation}
\label{eqn:definition_of_a}
\vec{a}\,^{(4)}(\vec{x})\equiv\left\{ a_{\ell_1 m_1}(r_1; \vec{x}),a_{\ell_2 m_2}(r_2; \vec{x}),a_{\ell_3 m_3}(r_3; \vec{x})\right\}.
\end{equation}

We then weight each $a_{\ell m}$ entering the sum dictated by $\mathcal{T}^{(4)}$ by its inverse power spectrum, which we denote $C_{\ell}^{-1}$; this in the end will cancel out power spectra of the $a_{\ell m}$ that enter after we compute their 4PCF and perform contractions. This weighting yields a new set, 
\begin{align}
    \vec{c}\,^{(4)}(\vec{x}) \equiv 
    \left\{ C^{-1}_{\ell_1} a_{\ell_1 m_1}(r_1; \vec{x}),C^{-1}_{\ell_2}a_{\ell_2 m_2}(r_2; \vec{x}),C^{-1}_{\ell_3}a_{\ell_3 m_3}(r_3; \vec{x})\right\}.
\end{align}
We then define
\begin{align}
\label{eqn: nabla4}
\nabla_{\ell_1m_1,r_1}\mathcal{T}^{(4)}[\vec{c}\,^{(4)}(\vec{x})]&\equiv\frac{\partial\mathcal{T}^{(4)}[\vec{c}\,^{(4)}(\vec{x})]}{\partial[a_{\ell_1 m_1}^*(r_1;\vec{x})]} - \left<\frac{\partial\mathcal{T}^{(4)}[\vec{c}\,^{(4)}(\vec{x})]}{\partial[a_{\ell_1 m_1}^*(r_1;\vec{x})]} \right>\\
&=\sum_{r_2<r_3}\sum_{\ell_2\ell_3}\sum_{m_2m_3}\zeta_{\ell_1\ell_2\ell_3}(r_1,r_2,r_3)\begin{pmatrix}
\ell_1 & \ell_2 & \ell_3 \\
m_1 & m_2 & m_3
\end{pmatrix}C_{\ell_2}^{-1}(r_2)C_{\ell_3}^{-1}(r_3)a^{\ast}_{\ell_2m_2}(r_2;\vec{x})a^{\ast}_{\ell_3m_3}(r_3;\vec{x})\nonumber\\
&\qquad- \left<\cdots \right>.\nonumber
\end{align}
Here, we must also subtract the expectation value in the first line, as we have an even number of $a_{\ell m}$ in this definition (as the second line shows) and so the expectation value of $\nabla \mathcal T^{(4)}$ would not otherwise vanish. Our notation in the last line simply means take the expectation value of the previous line. We subtract the expectation value because we want the perturbation to vanish on average; this will already occur naturally for $\nabla \mathcal{T}^{(3)}$, the quantity needed to realize a given 3PCF, as discussed in more detail in \cite{Slepian:2023gni}.

We now define
\begin{align}
\label{eqn:seven}
    \tilde{a}_{\ell m}(r_i; \vec{x}) \equiv a_{\ell_i m_i}(r_i; \vec{x}) + \frac{1}{2} \nabla_{\ell_i m_i, r_i} \mathcal{T}^{(3)}[\vec{c}\,^{(3)}(\vec{x})] + \frac{1}{3} \nabla_{\ell_i m_i, r_i} \mathcal{T}^{(4)}[\vec{c}\,^{(4)}(\vec{x})],
\end{align}
where $\vec{c}\,^{(3)}$ is the analogue of $\vec{c}\,^{(4)}$ but with only two $a_{\ell m}$ in it (see \citealt{Slepian:2023gni} Eq. 35). The superscipt-3 terms will give rise to a desired 3PCF, analogously to how the superscript-4 terms will give rise to a desired 4PCF. We define $\nabla_{\ell m}\mathcal{T}^{(3)}$ later, in Eq. (\ref{eqn: nabla3}), as we will explicitly use it in the next section.

We then correlate three such $\tilde{a}_{\ell m}$ around the same primary point, $\vec{x}$. After a calculation in \cite{Slepian:2023gni} \S3 which we do not reproduce here, this can be shown to yield a 4PCF about $\vec{x}$ with isotropic basis coefficients $\zeta_{\ell_1 \ell_2 \ell_3}$, as well as a 3PCF with desired coefficients in the two-argument isotropic basis functions (\textit{e.g.} \citealt{3pt_alg, encore}). 

We now briefly summarize why this occurs. When one forms the correlations of the different shells about $\vec{x}$, the original, unperturbed GRFs $a_{\ell m}$, contract with those inside the perturbation to give a non-vanishing value. The contractions eliminate the sums in Eq. (\ref{eqn:seven}) These contractions will also yield power spectra $C_{\ell_i}$, and these are then removed by the inverse power spectrum factors in our definition of $\vec{c}\,^{(4)}$. We are then left with $\zeta_{\ell_1 \ell_2 \ell_3}$. {\bf Essentially, the original GRFs act as ``carrier fields'' that carry with them, in the perturbation, the desired 3PCF and 4PCF correlations.}

This completes our review of how the \cite{Slepian:2023gni} algorithm produces, about a given primary, a density field with desired correlations. That work \textit{did not analyse} the impact of cross-correlations of points around different shells, and hence restricted itself to the geometry shown in Fig. \ref{fig:two_spheres_no_overlap}, where the sphere of radius $R_{\rm max}$ about a given primary could not overlap the sphere of radius $R_{\rm max}$ about another primary. The purpose of the current paper is to remove this restriction. In the following section, we show explicitly that these cross-correlations vanish.

\section{Cross-Correlating two different spheres}
\label{sec:cross-correlation}

\begin{figure}
\centering
\includegraphics[width = .65\linewidth]{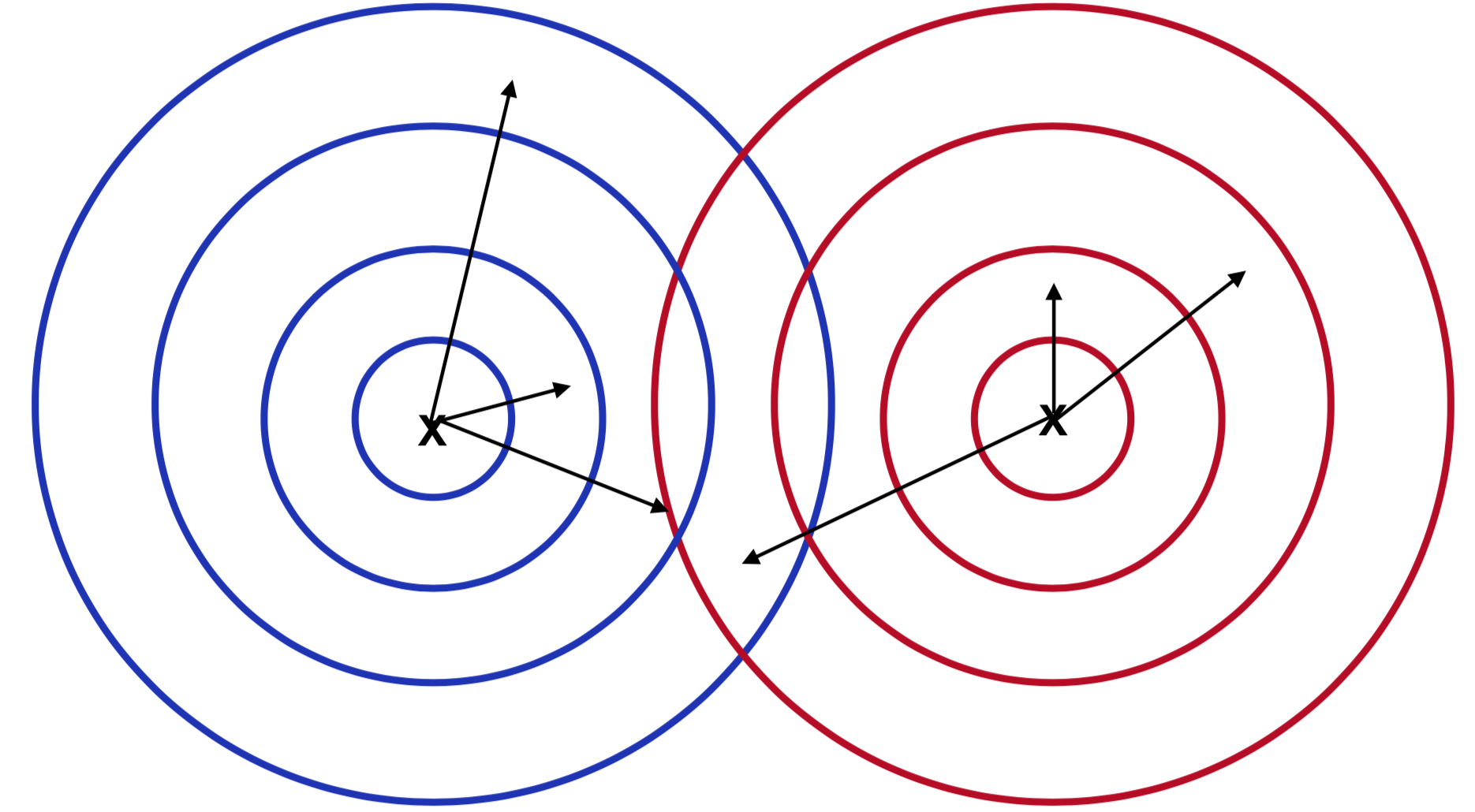}
\caption{Here we show the configuration under consideration in this paper, where the spheres around two primaries may overlap each other. However, because the GRFs on each shell and about each primary are independent, the red shells do not ``see'' the blue shells and vice versa.}
\label{fig:two_spheres_overlap}
\end{figure}

We consider here the situation represented in Fig.~\ref{fig:two_spheres_overlap}.

We first compute the 2PCF of the perturbed harmonic coefficients. We recall from \cite{Slepian:2023gni} that these are as in Eq. (\ref{eqn:seven}), with the additional definition
\begin{equation}
\label{eqn: nabla3}
\nabla_{\ell_1m_1,r_1}\mathcal{T}^{(3)}[\vec{c}^{\,(3)} (\vec{x})]=\sum_{r_2}\zeta_{\ell_1}(r_1,r_2)C_{\ell_1}^{-1}(r_2)a^{\ast}_{\ell_1m_1}(r_2;\vec{x}),
\end{equation}
and where $\nabla_{\ell_1m_1,r_1}\mathcal{T}^{(4)}[\vec{c}^{\,(4)}(\vec{x})]$ is given by Eq. (\ref{eqn: nabla4}).

The unperturbed $a_{\ell m}$ are all Gaussian Random Fields (GRFs), and they are by construction independent at each origin, $\vec{x}$, and on each spherical shell, $r_i$, and consequently have vanishing cross-correlation, except when evaluated around the same origin and on the same shell (which would be the auto-correlation). Mathematically, we then have:
\begin{equation}
\label{eqn: power_spectrum}
\expval{a_{\ell_1 m_1}(r_1;\vec{x}) \, a^{\ast}_{\ell_1^{\prime}m_1^{\prime}}(r_1^{\prime};\vec{x}^{\,\prime})}=C_{\ell_1}(r_1)\delta^{\mathrm{K}}_{\ell_1\ell_1^{\prime}}\delta^{\mathrm{K}}_{m_1m_1^{\prime}}\delta_{\mathrm{D}}^{[1]}(r_1-r_1^{\prime})\delta_{\mathrm{D}}^{[3]}(\vec{x}-\vec{x}^{\,\prime}),
\end{equation}
where $C_{\ell}$ is the power spectrum of the GRF used to produce the $a_{\ell m}$.

We now show that the cross-correlation of two perturbed $a_{\ell m}$ associated with the anisotropies in two different shells belonging to two different spheres vanishes too. Substituting Eqs.~\eqref{eqn: nabla3} and \eqref{eqn: nabla4}, we find
\begin{align}
\label{eqn:main}
\expval{\tilde{a}_{\ell_1 m_1}(r_1;\vec{x}) \,\tilde{a}^{\ast}_{\ell_1^{\prime}m_1^{\prime}}(r_1^{\prime};\vec{x}^{\,\prime})}=&\expval{a_{\ell_1 m_1}(r_1;\vec{x})a^{\ast}_{\ell_1^{\prime}m_1^{\prime}}(r_1^{\prime};\vec{x}^{\,\prime})}+\frac{1}{2}\sum_{r_2^{\prime}}\zeta_{\ell_1^{\prime}}(r_1^{\prime},r_2^{\prime})C_{\ell_1^{\prime}}^{-1}(r_2^{\prime})\expval{a_{\ell_1 m_1}(r_1;\vec{x})a^{\ast}_{\ell_1^{\prime}m_1^{\prime}}(r_2^{\prime};\vec{x}^{\,\prime})}\\
&+\frac{1}{2}\sum_{r_2}\zeta_{\ell_1}(r_1,r_2)C_{\ell_1}^{-1}(r_2)\expval{a^{\ast}_{\ell_1 m_1}(r_2;\vec{x})a_{\ell_1^{\prime}m_1^{\prime}}(r_1^{\prime};\vec{x}^{\,\prime})}\nonumber\\
&+\frac{1}{4}\sum_{r_2r_2^{\prime}}\zeta_{\ell_1}(r_1,r_2)\zeta_{\ell_1^{\prime}}(r_1^{\prime},r_2^{\prime})C_{\ell_1}^{-1}(r_2)C_{\ell_1^{\prime}}^{-1}(r_2^{\prime})\expval{a^{\ast}_{\ell_1m_1}(r_2;\vec{x})a^{\ast}_{\ell_1^{\prime}m_1^{\prime}}(r_2^{\prime};\vec{x}})\nonumber\\
&+\frac{1}{3}\sum_{r_2r_3}\sum_{\ell_2\ell_3}\sum_{m_2m_3}\zeta_{\ell_1\ell_2\ell_3}(r_1,r_2,r_3)\begin{pmatrix}
\ell_1 & \ell_2 & \ell_3 \\
m_1 & m_2 & m_3
\end{pmatrix}C_{\ell_2}^{-1}(r_2)C_{\ell_3}^{-1}(r_3)\nonumber\\
&\qquad\qquad\qquad\qquad\qquad\qquad\qquad\qquad\times\expval{a^*_{\ell_2m_2}(r_2;\vec{x})a^*_{\ell_3m_3}(r_3;\vec{x})a_{\ell_1^{\prime}m_1^{\prime}}(r_1^{\prime};\vec{x}^{\,\prime})}\nonumber\\
&+\frac{1}{3}\sum_{r_2^{\prime}r_3^{\prime}}\sum_{\ell_2^{\prime}\ell_3^{\prime}}\sum_{m_2^{\prime}m_3^{\prime}}\zeta_{\ell_1^{\prime}\ell_2^{\prime}\ell_3^{\prime}}(r_1^{\prime},r_2^{\prime},r_3^{\prime})\begin{pmatrix}
\ell_1^{\prime} & \ell_2^{\prime} & \ell_3^{\prime}\nonumber\\
m_1^{\prime} & m_2^{\prime} & m_3^{\prime}
\end{pmatrix}C_{\ell_2^{\prime}}^{-1}(r_2^{\prime})C_{\ell_3^{\prime}}^{-1}(r_3^{\prime})\nonumber\\
&\qquad\qquad\qquad\qquad\qquad\qquad\qquad\qquad\times\expval{a^*_{\ell_2^{\prime}m_2^{\prime}}(r_2^{\prime};\vec{x}^{\,\prime})a^*_{\ell_3^{\prime}m_3^{\prime}}(r_3^{\prime};\vec{x}^{\,\prime})a_{\ell_1 m_1}(r_1;\vec{x})}\nonumber\\
&+\frac{1}{6}\sum_{r_2r_3}\sum_{r_2^{\prime}}\sum_{\ell_2\ell_3}\sum_{m_2m_3}\zeta_{\ell_1\ell_2\ell_3}(r_1,r_2,r_3)\zeta_{\ell_1^{\prime}}(r_1^{\prime},r_2^{\prime})\begin{pmatrix}
\ell_1^{\prime} & \ell_2^{\prime} & \ell_3^{\prime} \\
m_1^{\prime} & m_2^{\prime} & m_3^{\prime}
\end{pmatrix}\nonumber\\
&\qquad\qquad\qquad\qquad\quad\times C_{\ell_1^{\prime}}^{-1}(r_2^{\prime})C_{\ell_2}^{-1}(r_2)C_{\ell_3}^{-1}(r_3)\expval{a_{\ell_2m_2}^*(r_2;\vec{x})a_{\ell_3m_3}^*(r_3;\vec{x})a_{\ell_1^{\prime}m_1^{\prime}}^*(r_2^{\prime};\vec{x}^{\,\prime})}\nonumber\\
&+\frac{1}{6}\sum_{r_2^{\prime}r_3^{\prime}}\sum_{r_2}\sum_{\ell_2^{\prime}\ell_3^{\prime}}\sum_{m_2^{\prime}m_3^{\prime}}\zeta_{\ell_1^{\prime}\ell_2^{\prime}\ell_3^{\prime}}(r_1^{\prime},r_2^{\prime},r_3^{\prime})\zeta_{\ell_1}(r_1,r_2)\begin{pmatrix}
\ell_1 & \ell_2 & \ell_3 \\
m_1 & m_2 & m_3
\end{pmatrix}\nonumber\\
&\qquad\qquad\qquad\qquad\quad\times C_{\ell_1}^{-1}(r_2)C_{\ell_2}^{-1}(r_2)C_{\ell_3^{\prime}}^{-1}(r_3^{\prime})\expval{a_{\ell_2^{\prime}m_2^{\prime}}^*(r_2^{\prime};\vec{x}^{\,\prime})a_{\ell_3^{\prime}m_3^{\prime}}^*(r_3^{\prime};\vec{x}^{\,\prime})a_{\ell_1m_1}^*(r_2;\vec{x})}\nonumber\\
&+\frac{1}{9}\sum_{\ell_2^{\prime}\ell_3^{\prime}}\sum_{m_2^{\prime}m_3^{\prime}}\sum_{r_2^{\prime}r_3^{\prime}}\sum_{\ell_2\ell_3}\sum_{m_2m_3}\sum_{r_2r_3}\zeta_{\ell_1\ell_2\ell_3}(r_1,r_2,r_3)\zeta_{\ell_1^{\prime}\ell_2^{\prime}\ell_3^{\prime}}(r_1^{\prime},r_2^{\prime},r_3^{\prime})\nonumber\\
&\qquad\qquad\qquad\qquad\times\begin{pmatrix}
\ell_1 & \ell_2 & \ell_3 \\
m_1 & m_2 & m_3
\end{pmatrix}\begin{pmatrix}
\ell_1^{\prime} & \ell_2^{\prime} & \ell_3^{\prime} \\
m_1^{\prime} & m_2^{\prime} & m_3^{\prime}
\end{pmatrix}C_{\ell_2}^{-1}(r_2)C_{\ell_3}^{-1}(r_3)C_{\ell_2^{\prime}}^{-1}(r_2^{\prime})C_{\ell_3^{\prime}}^{-1}(r_3^{\prime})\nonumber\\
&\qquad\qquad\qquad\qquad\qquad\qquad\qquad\quad\times\expval{ a^*_{\ell_2  m_2}(r_2;\vec{x})a^*_{\ell_3m_3}(r_3;\vec{x})a^*_{\ell_2^{\prime}m_3^{\prime}}(r_2^{\prime};\vec{x}^{\,\prime})a^*_{\ell_3^{\prime}m_3^{\prime}}(r_3^{\prime};\vec{x}^{\,\prime}) }.\nonumber
\end{align}
The terms proportional to $1/2$ and $1/4$ involve the unperturbed 2PCF which is zero for $\vec{x}\ne\vec{x}^{\,\prime}$. Moreover, the terms proportional to $1/3$ and $1/6$ vanish because they involve odd point correlation functions that are zero, since the unperturbed $a_{\ell m}$'s are Gaussian random fields. Let us then focus on the 4PCF in the last contribution (last line of Eq. \ref{eqn:main}). We use the theorem of \citet{isserlis1918formula}, also known as Wick's theorem:
\begin{equation}
\label{eqn:using_isserlis}
\begin{split}
&\langle a^*_{\ell_2m_2}(r_2;\vec{x})a^*_{\ell_3m_3}(r_3;\vec{x})a^*_{\ell_2^{\prime}m_3^{\prime}}(r_2^{\prime};\vec{x}^{\,\prime})a^*_{\ell_3^{\prime}m_3^{\prime}}(r_3^{\prime};\vec{x}^{\,\prime}) \rangle = \langle a^*_{\ell_2m_2}(r_2;\vec{x})a^*_{\ell_3m_3}(r_3;\vec{x})\rangle\langle a^*_{\ell_2^{\prime}m_3^{\prime}}(r_2^{\prime};\vec{x}^{\,\prime})a^*_{\ell_3^{\prime}m_3^{\prime}}(r_3^{\prime};\vec{x}^{\,\prime}) \rangle\\
&+\langle a^*_{\ell_2m_2}(r_2;\vec{x}) a^*_{\ell_2^{\prime}m_3^{\prime}}(r_2^{\prime};\vec{x}^{\,\prime})\rangle\langle a^*_{\ell_3m_3}(r_3;\vec{x})a^*_{\ell_3^{\prime}m_3^{\prime}}(r_3^{\prime};\vec{x}^{\,\prime})\rangle+\langle a^*_{\ell_2m_2}(r_2;\vec{x})a^*_{\ell_3^{\prime}m_3^{\prime}}(r_3^{\prime};\vec{x}^{\,\prime} )\rangle\langle a^*_{\ell_2^{\prime}m_3^{\prime}}(r_2^{\prime};\vec{x}^{\,\prime})a^*_{\ell_3m_3}(r_3;\vec{x}) \rangle.
\end{split}
\end{equation}
The terms in the last line simply vanish again because $\vec{x}\ne\vec{x}^{\,\prime}$. In the first line, we use that $r_2 \neq r_3$ and $r_2' \neq r_3'$---the fields around the same central point, but on different shells, are uncorrelated. Thus these expectation values all vanish: Eq. (\ref{eqn:using_isserlis}) equals zero, and this completes the proof that Eq. (\ref{eqn:main}) vanishes.

Now, because all fields are GRFs, the cross-correlations of shells around three different primaries, or around four different primaries, can be split into pairwise cross-correlations, by Wick's (Isserlis') theorem. we have just shown these pair correlations vanish; hence, this establishes that the corrrelations of triplets or quadruplets or beyond will also vanish. 

\section{Computational Cost}
\label{sec:computational_cost}
Consider the task of generating a density field that reproduces a given 3PCF and 4PCF up to a maximum multipole order of $\ell_{\text{max}} = 5$. To achieve this, we calculate the number of Gaussian random field (GRF) coefficients, $a_{\ell m}$, required for each spatial point $\vec{x}$. Assuming a discretization with $10$ radial bins for evaluating the 3PCF and 4PCF, each radial bin corresponds to a shell surrounding $\vec{x}$. On each shell, the spherical harmonic expansion includes $2\ell+1$ coefficients for each multipole $\ell$. Utilizing the conjugate symmetry of spherical harmonics, this count reduces to $\ell + 1$ independent coefficients per $\ell$. Therefore, the total number of coefficients on a single shell is given by the sum:
\begin{equation}
\sum_{\ell=0}^{\ell_{\text{max}}} (\ell + 1) = \frac{(\ell_{\text{max}} + 1)(\ell_{\text{max}} + 2)}{2}.
\end{equation}
To reconstruct both the 3PCF and 4PCF, we require separate sets of $a_{\ell m}$ coefficients for each, as illustrated in Eq.~\eqref{eqn: seven}. The size of these sets is identical, but the 4PCF involves more complicated combinations of coefficients than does the 3PCF, as evident from comparing Eqs.~\eqref{eqn: nabla3} and \eqref{eqn: nabla4}. Consequently, a factor of $2$ accounts for these two distinct sets. 

Including all \(10\) radial shells, the total number of \(a_{\ell m}\) coefficients required for \(\ell_{\text{max}} = 5\) is:
\begin{equation}
\frac{(\ell_{\text{max}} + 1)(\ell_{\text{max}} + 2)}{2} \times 2 \times N_{\text{shells}} = 6 \times 7 \times 10 = 420.
\end{equation}
This represents the number of GRF coefficients necessary for a single point, $\vec{x}$. This must be done around each grid point.

In total, then, the cost of our algorithm (to get a field with a given 3PCF and 4PCF) is to generate $N_{\rm GRF}$ Gaussian Random Fields, with
\begin{align}
    N_{\rm GRF} = N_{\rm grid}N_{\rm shells}(\ell_{\rm max} + 1)(\ell_{\rm max} + 2),
\end{align}
where $N_{\rm grid}$ is the number of grid points on which we wish to produce our density field.

A typical grid used to sample a survey, such as the Sloan Digital Sky Survey Baryon Oscillation Spectroscopic Survey (SDSS BOSS) CMASS or the DESI Year 1 Luminous Red Galaxy (LRG) sample volume, with reasonable resolution, might contain $4,096$ points in each dimension ($N_{\text{grid}}=4,096^3$). With the BOSS CMASS nominal volume of $8$ [Gpc/$h$]$^3$, this corresponds to a resolution of $0.1$ Mpc/$h$. We note that here the resolution has a different meaning than the usual---no N-body dynamics is being performed, so a discretisation is not producing inaccuracy in the positions of particles, and the force evaluations, at each time-step. Rather, here this is simply the precision to which an object's final position is given. Shifting objects' locations by of order $0.1$ Mpc/$h$ will have no effect on the measured clustering, since the NPCFs are binned to much fatter bins, of order 10 Mpc/$h$. Indeed, for many applications, even having a shift of order 1 Mpc/$h$ would be tolerable; this would allow using a much coarser grid, of order $512^3$.

For the 3PCF and 4PCF computations with $\ell_{\rm max} = 5$ and $N_{\text{shells}}=10$, we will need to draw a number of GRF values 
\begin{align}
N_{\rm GRF}=4,096^3\times420 \simeq 3 \times 10^{13}.
\end{align}
Each GRF value (\textit{i.e.} for a single $a_{\ell m}$ value on a single shell at a single point) can be drawn in of order thirty nanoseconds on a single core, resulting in a total wall-clock time of roughly 200 hours on a single core.\footnote{MacBook Pro 2018, 2.7 GHz Intel Core i7, 16 GB 2133 MHz RAM.} However, the computation is trivially parallelizable. Furthermore, the cost of doing an N-body simulation of such a volume is measured in millions of CPU hours. Even using a cheaper method like the Zeldovich approximation, as \texttt{EZmocks} \citep{ez-mocks} do, or several orders of perturbation theory on a grid, like \texttt{Grid-SPT} \citep{grid-spt}, is still far costlier.

Furthermore, as noted above, the resolution here simply means the precision of final object locations, and thus 1 Mpc/$h$ errors are probably acceptable for many applications; with the coarser, $512^3$ grid that permits, the time cost would drop by a factor of $\sim 10^3$, to $0.4$ hours on a  single core.

\section{Conclusion}
\label{sec:conclusion}
In this work, we have shown that the algorithm of \cite{Slepian:2023gni} to produce a density field with given, desired clustering around a \textit{specified} set of primary points, which are separated from each other by spheres of the maximum radius out to which clustering is measured, is more general than previously thought. Specifically, because the fields around each primary are statistically independent, one can eliminate the restriction that the points be separated as described above, as the independent fields around each primary do not ``see'' each other. We showed this explicitly at the level of correlations of pairs of points where each point is about a different primary. We then extended the argument to triplets, or quadruplets, or beyond, of points using the fact that the fields are Gaussian, so Wick's (Isserlis') theorem says all higher correlations can be reduced to pair correlations. 

With the restriction above removed, one can produce a grid of points where the clustering is as desired about \textit{all} of the grid points. In short, this yields a continuous (as represented on the machine) density field with a desired set of NPCFs.

We then calculated the number of GRFs one might require in a typical practical application, and using a reasonable number for the cost of realizing a GRF, we estimated the number of core hours the algorithm requires. The algorithm is trivially parallelizable, since each point and each field is independent, and thus the wall-clock time can be reduced significantly by using multiple cores. It is expected that this algorithm could produce density fields with desired NPCFs much faster than performing either an approximate or an N-body simulation.

\section*{Acknowledgments}
This work is funded by NASA grant number 80NSSC24M0021, for the Roman Galaxy Clustering Project Infrastructure Team.
Both authors also are grateful to the Slepian research group for useful conversations.

\section*{Data Availability}
All data relevant for this paper is given in the paper.


\clearpage

\bibliographystyle{mnras}
\bibliography{spheres.bib} 

\bsp	
\label{lastpage}
\end{document}